\documentclass{revtex4}
\usepackage{amsmath}
\usepackage{amsmath,amssymb,amsthm}
\usepackage{epsfig}
\usepackage{epstopdf}
\usepackage{graphicx}
\usepackage{verbatim}
\usepackage{hyperref}

\newcommand{\mapleplot}[4]{%
\begin{figure}[htb!]
\includegraphics[scale={#4}]{#1}
\caption{#2}
\label{#3}
\end{figure}}

\begin{document}

\title{Quasi-topological Reissner-Nordstr\"om Black Holes}
\author{W. G. Brenna$^{1}$\footnote{email address: wbrenna@uwaterloo.ca} and R. B. Mann$^{1}$\footnote{email address: rbmann@sciborg.uwaterloo.ca} }
\affiliation{$^1$Department of Physics \& Astronomy, University of Waterloo, 200 University Avenue West, Waterloo, Ontario, Canada, N2L 3G1}
\date{\today}

\begin{abstract}

We consider Reissner-Nordstr\"om solutions  in quasi-topological gravity, obtaining
exact solutions to the field equations yielding charged quasi-topological black holes.
We study their thermodynamic behaviour over a range of parameters that yield ghost-free
and stable space times.  We find that a sufficiently negative quasi-topological parameter 
can yield black holes with 2 horizons, even for zero charge.  We discuss the thermodynamic
stability for the class of solutions we obtain. We also describe the structure of exact charged
solutions to $k^{th}$ order quasi-topological gravity.
\end{abstract}

\maketitle

\section{Introduction}
\label{intro}

Gauge/gravity duality -- the notion  that gravitational dynamics in a given
dimensionality is dual to some other (nongravitational) field theory
of a lower dimensionality -- has been a subject of fruitful investigation in recent years.
Its chief instantiation has been via the AdS/CFT correspondence \cite{Maldacena}, and
investigations of this relationship from the perspective of Einstein gravity via the trace anomaly 
\cite{Sken} have indicated that the duality between central charges and couplings
on the non-gravitational side and the parameters on the gravitational side takes place
only for those  conformal field theories for which all the central charges are equal. Since this is
because there are not enough free parameters in  Einstein gravity, there has been motivation
to  broaden the universality class of dual field theories to more general theories of gravity  
containing more free parameters. Examples are Lovelock theory \cite{Lov} and various forms
of quasi-topological gravity \cite{Oliva,Myers1,Dehghani:2011vu}, whose additional central
charges have recently been investigated holographically \cite{Myers2,Myers3}.

We are concerned here with cubic quasi-topological gravity coupled to electromagnetism in $D$ dimensions.
Such higher-order curvature terms modify the gravitational action in a way
that may exhibit useful dualities, in a manner similar to the   AdS/CFT correspondence \cite{Maldacena}.
For example, requiring the dual gauge theory to respect causality
 (by considering graviton fluctuations in the gravity theory, which probe the bulk geometry in the presence of a black hole)
yields constraints on the coupling constants of the gravity theory so that the dual theory
does not support superluminal signals.  In  Lovelock gravity
these causality constraints precisely match those
arising from requiring positive energy fluxes \cite{Lg2, Lg3}.  However, this matching does not appear in general, in particular
for cases where the gravitational equations of motion are not second order \cite{Hof}.  
Cubic quasi-topological gravity is of interest insofar as there are three constraints that arise from requiring positive energy fluxes,
which determine the three coupling constants. No evidence for
causality violation has been observed  when the  curvature-cubed couplings are consistent with these constraints \cite{Myers2}. 

Quasi-topological gravity  is akin to Lovelock gravity insofar as  
the field equations are second order differential equations with respect to the 
metric, though it produces nontrivial modifications to the action in dimensions where Lovelock gravity is
a topological invariant \cite{Oliva}. However,
these additional dimensions come with the caveat that the field equations are
second order only when the solution is spherically symmetric. The nomenclature
``quasi-topological''  is employed because in the dimensions for which 
its terms do not normally affect the action, breaking spherical symmetry 
causes it to contribute nontrivially to the action (and therefore, quasi-topological terms
are not true topological invariants), yielding $3^{rd}$ order differential equations. 
Despite this,  one of the more intriguing features of quasi-topological gravity is that the linearized equations describing gravitons propagating in an AdS vacuum match precisely the second-order equations of Einstein gravity. It is therefore possible to find
stable vacua in the theory that are free of ghosts \cite{Myers1}.  
  
 Here we seek
  charged Reissner-Nordstr\"om (RN) black hole solutions to the field equations of quasi-topological gravity, 
and analyze their behaviour.   RN black holes are significant in Einsteinian gravity, featuring a characteristic
dual event horizon because of their charge.  As in the Lifshitz case \cite{Brenna:2011gp},
all the desirable properties of quasi-topological gravity are 
preserved upon coupling to a Maxwell field, and so it is reasonable to explore the physics of charged solutions.
Indeed, our motivation for the study of these black holes is the utility of their charge
parameter in the context of gauge-gravity duality.
The goal in a search for dual theories is the development of a ``dictionary'' of
parameters, which allows one to evaluate quantities in one theory and map them
to another. Work has been performed which hints towards dualities between higher curvature
theories and exotic condensed matter behaviour, such as the proposed relationship
between Gauss-Bonnet gravity and non-Fermi liquids \cite{pourhasan}.
In addition, studies have been performed to examine relationships between
uncharged quasi-topological gravity with probe-limit fields and holographic superconductors \cite{ling}.

It would be particularly enticing to perform similar work in quasi-topological gravity
for a number of reasons.
First, the third-order quasi-topological terms become active (nontrivial) in five dimensions,
allowing us to hypothesize a dual between a five dimensional quasi-topological theory
and a four dimensional physical theory, with the bonus of a tuneable quasi-topological parameter.
In addition, the seemingly unnatural constraint of spherical symmetry in quasi-topological gravity
becomes unimportant, because the vast majority of metrics we understand well are spherically symmetric,
so a search for spherically symmetric dual behaviour is likely to be equally profitable to a more general search.

RN black holes feature a charge parameter in addition to their mass, which may be required
to reproduce a constant low-temperature Sommerfeld ratio (as in Gauss-Bonnet gravity \cite{pourhasan}).
The Sommerfeld ratio's behaviour is a fundamental property of non-Fermi liquid systems,
and recovering a constant Sommerfeld ratio at low temperature is a start to the development
of a holographic duality.
We will explore the quasi-topological behaviour of charged black holes in order
to build a foundation for their use as quasi-topological gravitational duals.

In this paper we obtain exact solutions for quasi-topological black holes with a U(1) 
electromagnetic charge  in asymptotically anti de Sitter (AdS), de Sitter (dS), and flat spacetimes.
We shall choose the parameters of quasi-topological gravity such that the criteria for 
stability and ghost-freedom are satisfied. 
Of interest is the effect of the quasi-topological parameter $\mu$ on  black hole thermodynamics.
After considering how the temperature-entropy relationship is modified, we shall examine the specific heat of the charged black holes,
to see how  $\mu$ allows us to explore a larger range of possibilities for
phase-transitions than in the case of Gauss-Bonnet gravity alone.

We find that a sufficiently negative quasi-topological parameter is able to produce uncharged spherical black hole
solutions with two event horizons in asymptotically AdS, dS, and flat spacetimes, and that such negative corrections
also push black hole solutions towards extremality when compared with Einsteinian solutions of the same
outer horizon radius.
We also find that the quasi-topological parameter provides greater control over the coefficients of a
fifth order polynomial which governs phase transitions (as the denominator of the specific heat).
Thus, it may be possible to tune the quasi-topological parameters to induce and remove phase transitions which
do not occur in the Einsteinian case.

\section{$3^{rd}$-order Quasi-topological Gravity}
\label{adssol}

We consider quasi-topological additions corresponding to $3^{rd}$-order curvature corrections 
to Einstein-Maxwell-Gauss-Bonnet gravity that maintain second-order field equations with respect 
to the metric under conditions of spherical symmetry.  
We use the action
\begin{equation}
I = \int d^{D} x \sqrt{-g} \left( - 2 \Lambda + \mathcal{L}_1 + \frac{\lambda L^2}{(D-3)(D-4)}\mathcal{L}_2 + \frac{7 \mu L^4}{4}\mathcal{L}_3 - \frac{1}{4}F_{\mu \nu}F^{\mu \nu}   \right)
\label{action}
\end{equation}
where $D$ is the number of dimensions, 
$F_{\mu \nu} = \partial_{[\mu}A_{\nu]}$, $\mu$ and $\lambda$ are the correction terms' coefficients, 
$\mathcal{L}_1 = R$ is the Ricci scalar, 
$\mathcal{L}_2 = R_{\mu \nu \gamma \delta} R^{\mu \nu \gamma \delta} - 4 R_{\mu \nu} R^{\mu \nu} + R^2$ 
is the Gauss-Bonnet Lagrangian, and $\mathcal{L}_3$ is the $3^{rd}$ order quasi-topological gravity term.
This  term has the form 
\begin{align}
\mathcal{L}_3 &= {{{R_\mu}^\nu}_\alpha}^\beta {{{R_\nu}^\tau}_\beta}^\sigma {{{R_\tau}^\mu}_\sigma}^\alpha + \frac{1}{(2D - 3)(D - 4)} \left( \frac{3(3D - 8)}{8} R_{\mu \alpha \nu \beta} R^{\mu \alpha \nu \beta} R \right. \nonumber \\
              & \quad - 3(D-2) R_{\mu \alpha \nu \beta} {R^{\mu \alpha \nu}}_\tau R^{\beta \tau} + 3D\cdot R_{\mu \alpha \nu \beta} R^{\mu \nu} R^{\alpha \beta} \nonumber \\
              & \quad \left. + 6(D-2) {R_\mu}^\alpha {R_\alpha}^\nu {R_\nu}^\mu - \frac{3(3D-4)}{2} {R_\mu}^\alpha {R_\alpha}^\mu R + \frac{3D}{8} R^3 \right)
\label{quasitop}
\end{align}
and is only effective in dimensions greater than $D>4$, becoming trivial 
in 6 dimensions \cite{Myers}.

The metric ansatz we employ is
\begin{equation}\label{metric}
ds^2 = -\frac{r^2}{L^2} f(r) dt^2 + \frac{L^2 dr^2}{r^2 g(r)} + \frac{ r^2 }{ L^2 } d\Omega_k^2
\end{equation}
where $d\Omega_k^2$ is the metric of a constant curvature hypersurface
\begin{equation}
d\Omega_k^2 =d{\theta_1}^2 + k^{-1}\sin^2 {\left(\sqrt{k} \theta_1\right)} \left( d{\theta_2}^2 + \displaystyle\sum\limits_{i=3}^{3} \displaystyle\prod\limits_{j=2}^{i-1} \sin^2{\theta_j } d\theta_i^2 \right)
\end{equation}
The parameter $k=-1,0,1$, corresponding to hyperbolic, flat, and spherical geometries, respectively. We parametrize the Maxwell gauge field with a function $h(r)$ 
\begin{equation}\label{gfield}
A_t = q \frac{r}{L} h(r) 
\end{equation}
and additionally introduce the notation $j(r) = \frac{dh(r)}{dr}$ for convenience.

The field equations that follow from the action are
\begin{align}
& \Lambda L^2 r^6 + \left( 6 \right) r^6 g - 6 \lambda r^6 g^2 + 6 \lambda r^4 L^2 k g - 3 r^4 L^2 k - \left( 6 \right) \mu r^6 g^3 + \left( 9 \right) \nonumber \\
& \mu r^4 L^2 k g^2 - 3 \mu L^6 k^3 + g (\ln{f})^{'} \left( \frac{3}{2} r^7 - 3 \right. \lambda r^7 g + 3 \lambda r^5 L^2 k - \frac{9}{2} \mu r^7 g^2 \nonumber \\
& + \left. 9 \mu r^5 g L^2 k - \frac{9}{2} \mu r^3 L^4 k^2 \right) = - \frac{q^2 r^6}{4 f}\left[ g \left(r h^{'} +  h \right)^2  \right] \label{fieldequations_initial}\\
& \left( 3 r^4 \left[ - \frac{\Lambda}{6} L^2%
- \kappa + \lambda \kappa^2 + \mu \kappa^3 \right] \right)^{\prime}%
= \frac{q^2 r^3}{2 f}\left[ g \left(r h^{'} +  h \right)^2  \right]
\label{fieldequations_initial2}\\
&2r^{2}h^{\prime \prime }- r\left[ (\ln f)^{\prime }-(\ln%
g)^{\prime }\right]( rh^{\prime }+h) +10 rh^{\prime }+6h=0
\label{fieldequations_final}
\end{align}
using (\ref{metric},\ref{gfield}),  where $\kappa = \left(g - \frac{L^2}{r^2} k \right)$.  These equations simplify considerably if we substitute  $f(r) = N^2(r) g(r)$, yielding
\begin{eqnarray}
(-1+2 \lambda \kappa+ 3\mu \kappa^2) N^\prime &=& 0 \label{finalN}\\
\left( 3 r^4 \left[ - \frac{\Lambda}{6} L^2%
- \kappa + \lambda \kappa^2 + \mu \kappa^3 \right] \right)^{\prime}%
&=& \frac{q^2 r^3}{2}\left[  \left(\frac{\left(r h \right)^\prime}{N}\right)^2  \right]
\label{finalg}\\
\left( \frac{r^3}{N} \left(r h \right)^\prime  \right)^\prime  &=&0
\label{finalh}
\end{eqnarray}

\section{Black Hole Solutions}
\label{bhsol}

We consider here the exact solutions to the field equations in five dimensions with the addition of the 
quasi-topological term.  It is clear that $N=1$ solves one of the field equations.  We first begin with the $h=0$
case.
 
Setting $h=0$, it is clear that we have one equation to solve:
\begin{equation}\label{cub1}
3 r^4 \left[ - \frac{\Lambda}{6} L^2%
- \kappa + \lambda \kappa^2 + \mu \kappa^3 \right]  = M
\end{equation}
which is cubic in $\kappa$.  There are three branches to the solution, only one of which reduces to the Reissner-Nordstr\"om solution when $\lambda=\mu=0$.  
The principal solution (which does not necessarily reduce to the Reissner-Nordstr\"om solution) is
\begin{equation}\label{metQ0}
g(r) = \frac{k L^2}{r^2} - \frac{\lambda}{3 \mu} +  \frac{1}{12 \mu } \left[ \left( %
\sqrt{ \Gamma + J^2(r) } + J(r) \right)^{\frac{1}{3}} - \left( \sqrt{ \Gamma + J^2(r)} - J(r) \right)^{\frac{1}{3}} \right]
\end{equation}
where we define  
\begin{align*}
\Gamma &= - \left( 16\left( 3 \mu + \lambda^2 \right) \right)^3 \\
J(r)   &= 36  \left( -4 \mu^2 \Lambda L^2 - 8 \frac{M \mu^2}{r^4} + 8 \mu \lambda  + \frac{16}{9} \lambda^3 \right)  
\end{align*}
and $M$ is a constant of integration.   
This solution matches the form recently obtained for the 7-dimensional $3^{rd}$ order Lovelock case \cite{Pour}, since the field equations  
are the same apart from powers of $r$ that arise due to this difference in dimensionality.
 
The charged solutions are obtained by solving (\ref{finalh})  
\begin{equation}\label{Asol}
A = Q  \frac{dt}{2\sqrt{2} L r^2}
\end{equation}
in turn yielding 
\begin{align}
\label{emresult}
F &= - \frac{Q}{r^3 \cdot L \sqrt{2}}\,  dr \wedge dt
\end{align}
The solution of (\ref{finalg}) is then given by
\begin{equation}
 - \frac{\Lambda}{6} L^2 - \left( g - \frac{L^2}{r^2} k \right) + \lambda \left( g - \frac{L^2}{r^2} k \right)^2 + \mu \left( g - \frac{L^2}{r^2} k \right)^3%
 = \frac{M}{3 r^4} - \frac{Q^2}{24 r^6} 
 \label{eqnforg}
\end{equation}
where $Q$ and $M$ are constants of integration and $\Lambda$ is the cosmological constant.  This is
the $3^{rd}$-order quasi-topological generalization of the Reissner-Nordstr\"om solution.

The principal branch of the solution is the
same as that of (\ref{metQ0}), but with
\begin{equation}\label{QJ}
J(r) =  36  \left( -4 \mu^2 \Lambda L^2 - 8 \frac{M \mu^2}{r^4} + 2\frac{Q^2}{r^6} + 8 \mu \lambda  + \frac{16}{9} \lambda^3 \right) 
\end{equation}
To ensure that the principal solution of the cubic is real, we find that the inequality
\begin{equation}
\label{restriction}
\mu < -\frac{\lambda^2}{3}
\end{equation}
must be satisfied.

Turning to look at all branches,  we rewrite eq. (\ref{cub1}) for $g(r)$ as
\begin{equation}\label{cub2}
W[\kappa] \equiv
  \left[\epsilon
- \kappa + \lambda \kappa^2 + \mu \kappa^3 \right]  =  \frac{M}{3 r^4} - \frac{Q^2 }{24 r^6}
\end{equation}
where $\epsilon = -\Lambda L^2/6$ is respectively equal to $1,-1,0$ for asymptotically AdS, dS, and  flat solutions.  
The Kretschmann scalar diverges if the condition
\begin{equation*}
\partial_r \kappa = \frac{d}{dr} W[\kappa] \left[ \frac{\partial W[\kappa]}{\partial \kappa}  \right]^{-1}
\end{equation*}
is satisfied.  The denominator corresponds to extremal values of $W[\kappa]$ with respect to $\kappa$. 
In the charged case, the numerator can also equal zero for
\begin{equation*}
Q^2 = \frac{8}{3 } r_{ns}^2 M .
\end{equation*}
where $r_{ns}$ is the radius corresponding to the solution of the extremal value of $W[\kappa]$.

For simplicity we do not consider the case when the numerator reaches zero (this could result in a $0/0$ solution),  
and broadly disregard solutions
where $W[\kappa]$ reaches extremal values.
Taking the derivative of equation (\ref{cub2}), we find that extremal values occur when
\begin{equation}
g(r_{ns}) = k \frac{L^2}{ {r_{ns}}^2} - \frac{\lambda}{3 \mu} \pm \sqrt{ \left(\frac{\lambda}{3 \mu} \right)^2 + \frac{1}{3 \mu}  }.
\end{equation}
To ensure that this never happens, we can force the discriminant to be negative (given that $g(r)$ is always real as per restrictions above) yielding
\begin{equation*}
\mu < - \frac{\lambda^2}{3} .
\end{equation*}
We therefore see that by satisfying equation (\ref{restriction}) we obtain
meaningful roots.
The value of $\Lambda$ is all that remains to be chosen.
For simplicity, we choose between three values of $\Lambda$;
one for AdS, one for dS, and one for flat-space black hole solutions.

In what follows, the values of $\lambda$ and $\mu$ shall be chosen to lie within the region where our theory has stable AdS vacua
without ghost gravitons, and where nonsingular planar black hole solutions exist \cite{Myers}.
Unless otherwise stated, we shall use $\lambda=0.04$ and $\mu = -0.001$.   Note that for $\lambda = 0.04$  we remain
within this region provided  $\mu \leqslant -0.000408$.  The solution (\ref{metQ0},\ref{QJ}) is the only real solution within this region of $\lambda$ and $\mu$.  We note that as $\lambda,\mu \rightarrow 0$, this solution does not approach the Einsteinian black hole.
Taking $\mu \rightarrow 0$, as discussed above, will lead to imaginary components in the function $g(r)$ and the solution is deemed unphysical.  Indeed,
examining the roots of the equation (\ref{eqnforg}), one can observe that the real root travels to $\pm \infty$ as the parameters
$\lambda,\mu \rightarrow 0$ (depending on the direction from which 0 is approached).
Nonetheless, though the branch we select does not ultimately converge to the Einsteinian branch in this limit, we shall see that
the thermodynamic behaviour does become very similar
for $\mu=-0.001$ and a small $\lambda$,
and so we can make reasonable conclusions about the effects of a quasi-topological parameter on black hole thermodynamics.
In addition, the other two branches of the field equations only produce nonsingular solutions in regions where the AdS vacuum may be ghosty \cite{Myers}.
Though nakedly singular solutions have been shown to have some observable characteristics \cite{virbhadra},
we will choose to study black hole thermodynamics in this paper because they are more well-understood as conjectured duals
in the gauge/gravity correspondence.


\subsubsection{Asymptotically anti de Sitter solutions}

We constrain $\Lambda$ for these solutions such that they admit the limit $\lim_{r\to \infty}g(r) = 1$, using the field equation
(\ref{eqnforg}):
\begin{equation*}
\Lambda =- \frac{6 \left(1- \mu - \lambda  \right)}{L^2}
\end{equation*}
This restriction is introduced simply because it is convenient to compare metric functions that converge to the same value;
the effects of the higher curvature terms can be more easily seen.
When we subsequently discuss thermodynamics, we shall see that this definition of the cosmological constant will artificially
affect the temperature/entropy relationship, and so it will become more convenient to compare solutions with the
same cosmological constant. 

We are then able to obtain Reissner-Nordstr\"om solutions for a given black hole radius.
These solutions are asymptotically anti de Sitter provided the cosmological constant stays negative, that is, $\mu + \lambda < 1$.
In order to meaningfully compare results for various cases, we consider black holes whose event horizons
have the same radius, which in turn implies that they have different mass.

Figure \ref{reisnordcomp} shows three solutions with an event horizon of $r_+ = 1.0$ for $L=1$ ($L=1$ is assumed throughout the paper): $Q=0$ (solid), $Q=12$ (dashed), and $Q=6$ (dotted).
It turns out that $Q=6$ is near-extremal (the inset of Figure \ref{reisnordcomp} indicates 2 horizons),
and so Reissner-Nordstr\"om solutions with significantly greater charge than this
do not exist for the given event horizon radius.  For  $Q=12$ we plot an extremal solution since 
we cannot obtain a  $r_+ = 1.0$ black hole for this case.  The respective masses for $Q=0,6,12$ are $M=2.883000024$,
 $M=7.383000024$ and  $M=18.483$; just beyond the near-extremal $Q=6$ case  graphical
methods yield an approximate extremal mass of $M \approx 7.3315$.


\mapleplot{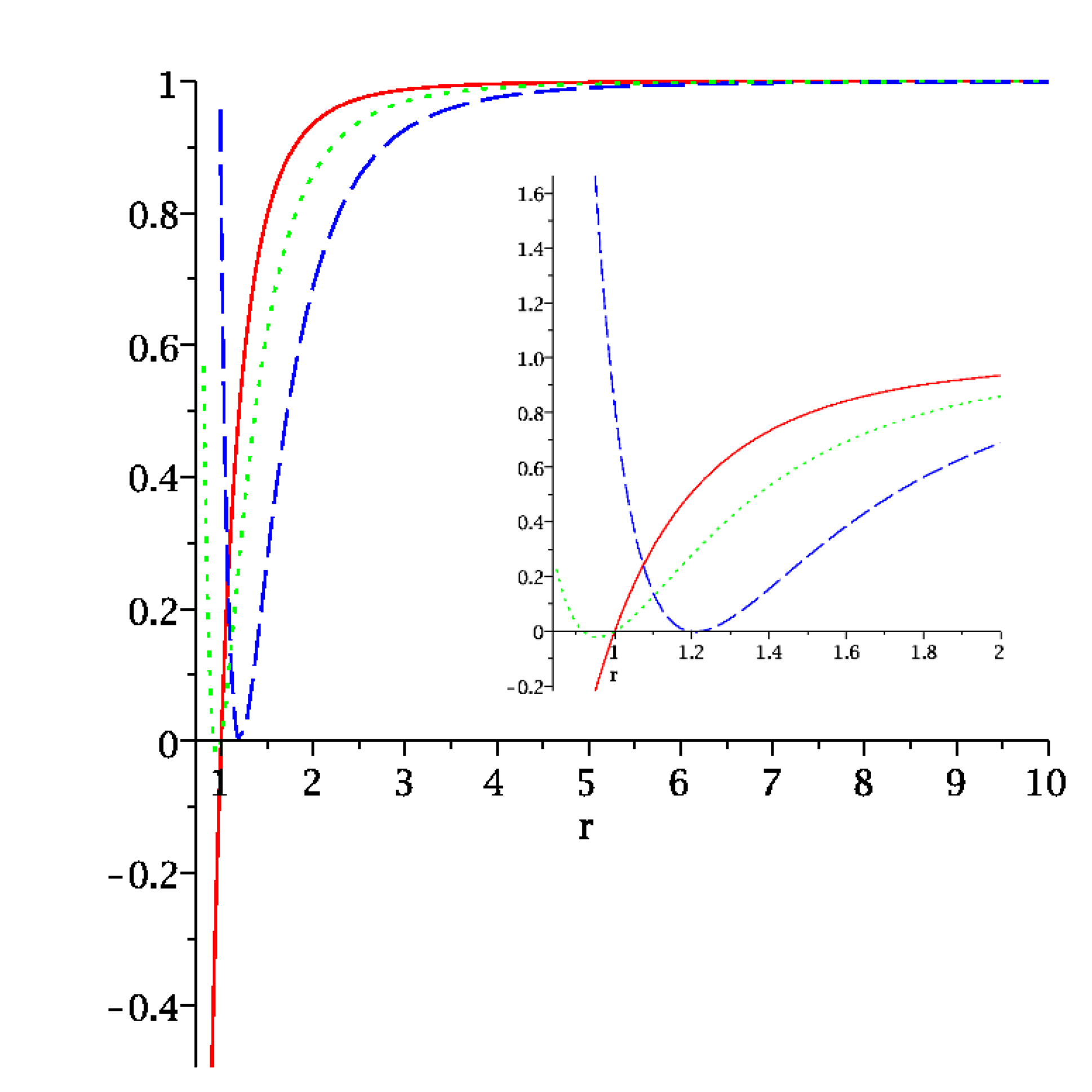}{Comparison of Reissner-Nordstr\"om and uncharged Black Hole Solutions' metric functions versus
radial coordinate for $k=0$.  
The solutions are $Q=0$ (solid), $Q=6$ (dotted), and $Q=12$ (dashed). The inset is zoomed in to enhance the near-horizon behaviour of $g(r)$. 
}{reisnordcomp}{0.4}

For given values of $Q$ and one of the roots of $g(r )$, the value of $M$ can be obtained from eq (\ref{eqnforg}). To ensure
that the root of $g(r )$ is the outer horizon $r_+$, we used an 
iterative method to determine the given $M$ parameter values which corresponded to
particular roots at $r=r_+$. To ensure accuracy in locating the roots,  we required that the iteration
algorithm not cease until convergence to within $10^{-10}$ of our desired $r_+$ was attained. 
For black holes without a quasi-topological parameter ($\mu=0$), we used analytic solutions that
correspond to the value of $M$ for any given rightmost root $r_+$, as previously done in the chargeless case \cite{Boulware1985}.
These solutions are listed in Appendix \ref{asolnomu}.

Figures \ref{rnr2kn1}--\ref{rnr2k1} respectively illustrate solutions for $k=-1,0,1$ of the metric
functions of  black holes with outer radius $r_+ = 2.0$, for
$Q$ values of $0$, $6$, and $12$.  
\mapleplot{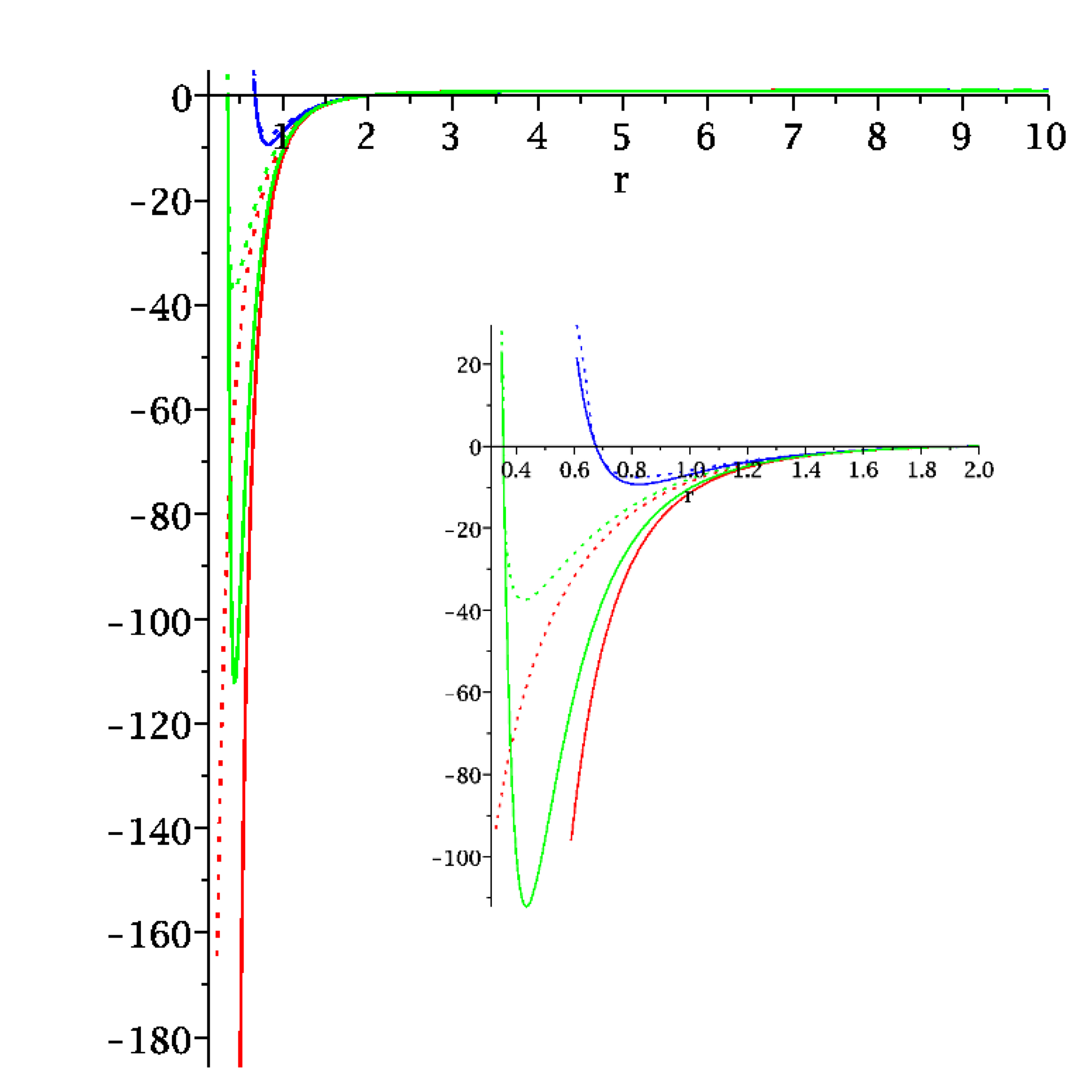}{Reissner-Nordstr\"om Black Hole Solutions for $k=-1$, $r_+ = 2.0$, where $Q=0,6,12$ are in red, green, and blue, respectively.
The dotted solutions are quasi-topological, and the solid solutions are Einsteinian. }{rnr2kn1}{0.4}
\mapleplot{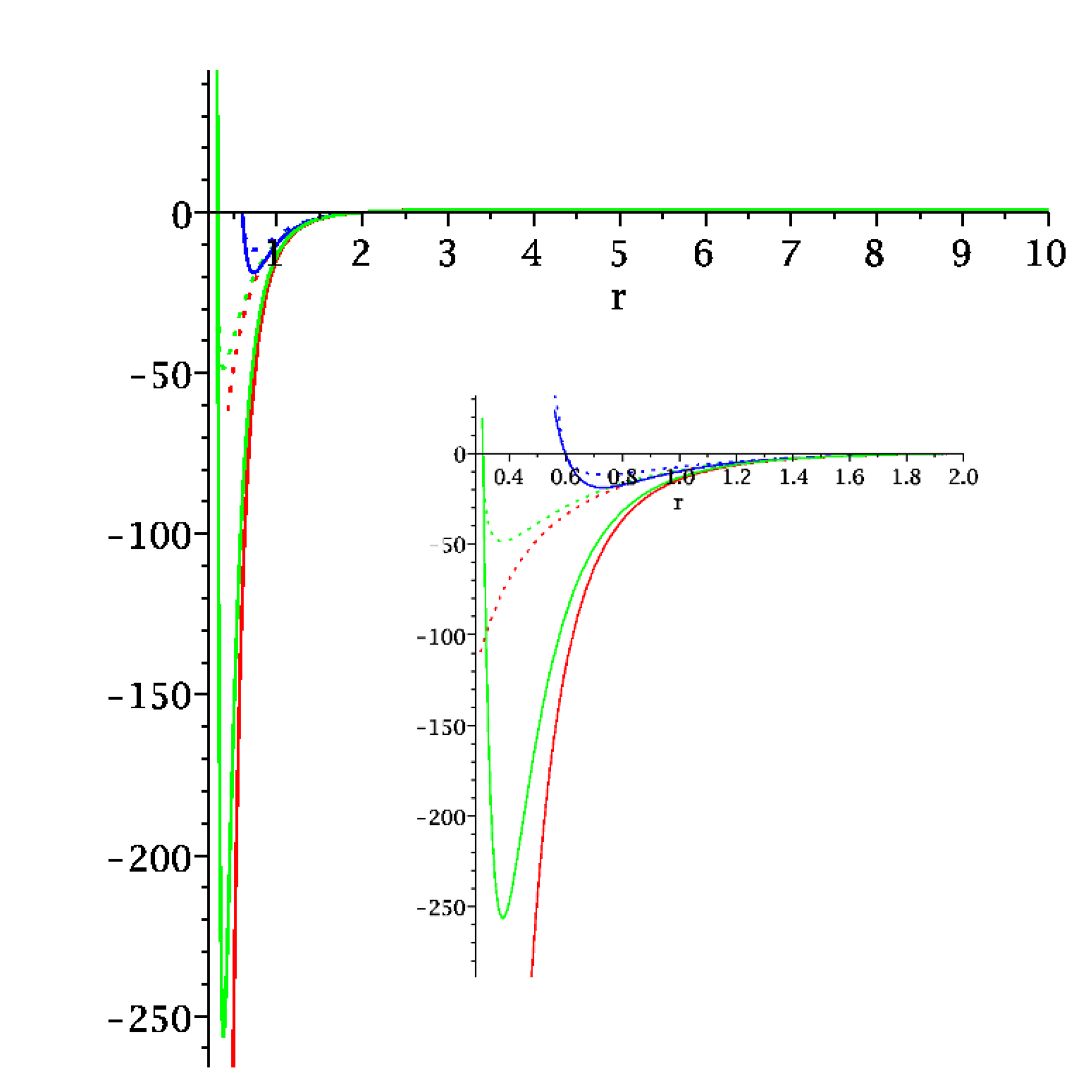}{Reissner-Nordstr\"om Black Hole Solutions for $k=0$, $r_+ = 2.0$, where $Q=0,6,12$ are in red, green, and blue, respectively.
The dotted solutions are quasi-topological, and the solid solutions are Einsteinian. }{rnr2k0}{0.4}
\mapleplot{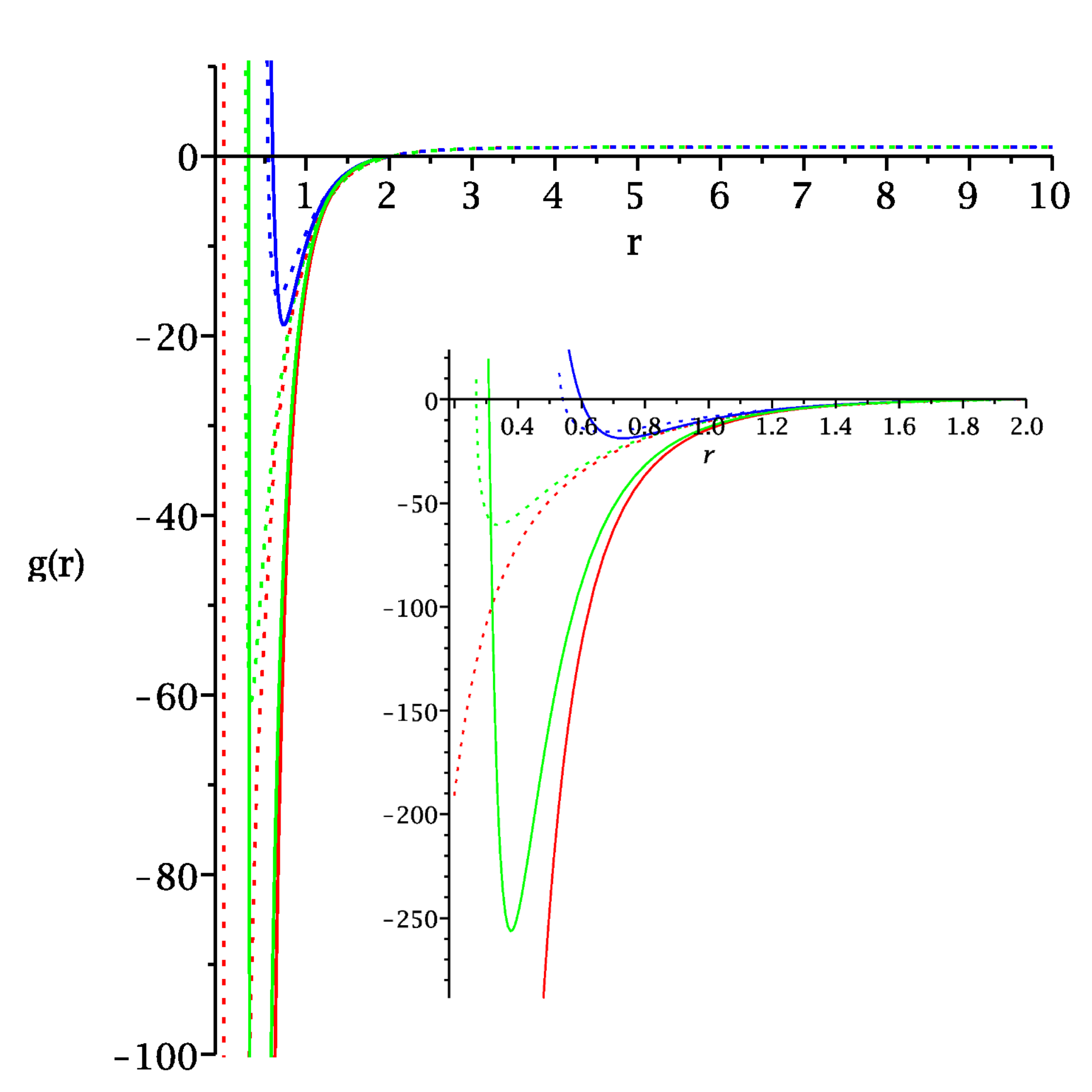}{Reissner-Nordstr\"om Black Hole Solutions for $k=1$, $r_+ = 2.0$, where $Q=0,6,12$ are in red, green, and blue, respectively.
The dotted solutions are quasi-topological, and the solid solutions are Einsteinian. }{rnr2k1}{0.4}
We find that the functions $f(r) = g(r)$ possess two linear zeroes (and so the metric has
two event horizons) in the charged cases, and that they properly asymptote to $1$  (though this is difficult to see due to the scale of the graph).

The inlay plots show that the quasi-topological solution (with $\lambda = 0.04$ and $\mu = -0.001$)
modifies the behaviour in a counterintuitive manner - it does not become ``more extremal'' because the inner horizon does not converge towards the event horizon;
instead for $k=1$ the inner horizon actually moves to a smaller radial value, while for $k=0$ and $k=-1$ the inner horizon does not appear to change significantly
from the Einsteinian case.
The higher curvature corrections also act to decrease the magnitude of the negative spike in $g(r)$. 

Interestingly, for $k=1$ the quasi-topological solution with $Q=0$  induces two event horizons (which can just be seen in Figure \ref{rnr2k1}).
In this sense quasi-topological gravity induces charge-like effects into the space-time structure, a phenomenon that
we shall see is also present in asymptotically dS and flat spacetimes.

\subsubsection{Asymptotically de Sitter solutions}

Requiring the metric functions $g(r)$ to tend to $-1$ as $r  \rightarrow \infty$ yields from (\ref{eqnforg})
\begin{equation*}
\Lambda = \frac{6 \left( 1 - \mu + \lambda  \right)}{L^2}
\end{equation*}
and we shall adopt this value of $\Lambda$ for this section.

We can see that for $\lambda - \mu > -1$, the spacetime is de Sitter.
The function $g(r)$ is plotted in Figure \ref{desitter}, for a mass parameter of $0.7$.
In this case, the differently charged black holes do have different event horizon radii, 
though they do not differ greatly; the event horizons are all between $r_+ = 0.8$ and $r_+ = 0.9$.
We can see that increased charge results in the appearance of naked
singularities, and zero charge causes a negative asymptote (i.e. the solution then has only one horizon).

Note that with our requirement that $g(r)$ tends to $-1$, we also need the presence of a cosmological
horizon at some $r = r_C > r_+$, which is present in the solution of equation (\ref{metQ0}) only if $k=1$; otherwise
this horizon will not be present.  
In this case, the negative quasi-topological parameter modifies the solutions in the same
direction as the positive Gauss-Bonnet parameter.
These parameters also act oppositely to the AdS case - they now do induce ``more extremal'' behaviour in
solutions. Sub-extremal solutions in Einstein gravity move beyond extremality for even the small values
of $\lambda$ and $\mu$ we consider; specifically,  $Q=0.6$ is not extremal in the Einsteinian case, but is seen 
from fig. \ref{desitter} to be nakedly singular
for both Gauss-Bonnet and quasi-topological solutions, where only one horizon exists.
 
Recall that for a value of $\lambda = 0.04$, the condition on $\mu$ producing a ghost-free AdS vacuum is
$\mu < -0.000408$.  We find in this asymptotically dS context that for values of $\mu$ more negative than this we obtain solutions with three horizons
even for $Q=0$. In this manner the gravitational effects of charge thus are induced by the higher curvature corrections
in quasi-topological gravity.  However a sufficiently negative value of $\mu$ yields solutions
with a naked singularity and one cosmological horizon. For example a value of $\mu \approx -0.011$ causes both the $Q=0.2$ and $Q=0$ solutions to move past extremality, yielding singularities surrounded by
a cosmological horizon only.     We note that in this example, $\mu < -0.001$ with $\lambda = 0.04$ does not in the asymptotically AdS context  guarantee positivity of energy flux in the dual CFT \cite{Myers2}. In the asymptotically dS context we are considering in this section we find that violating this bound on $\mu$  can yield nakedly singular solutions.

Attempting to produce for $Q=0$ a space-time with only 2 horizons
entails increasing $\mu$ above the region where our solution is real. Hence  the ``charge''-like behaviour induced by the quasi-topological term is a result of our branch not reducing to the Einsteinian branch.

\mapleplot{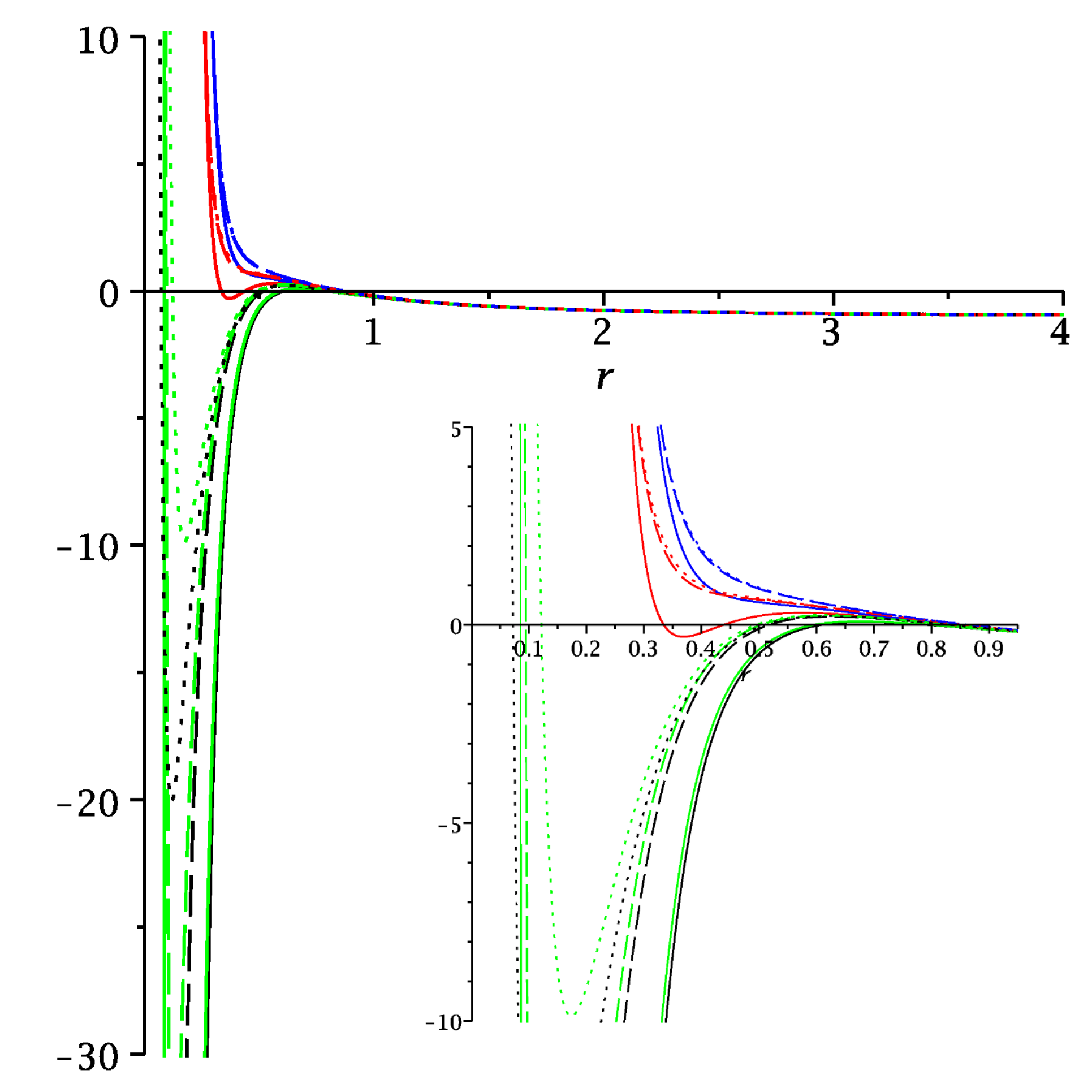}{Solutions in de Sitter space for $k=1$, where $Q=0,0.2,0.6,0.7$ are in black, green, red, and blue, respectively. Here $\lambda = 0.04$ and $\mu = -0.001$. Quasi-topological gravity is dotted, Gauss-Bonnet is dashed, and Einsteinian is solid.}{desitter}{0.4}

\subsubsection{Asymptotically flat solutions}

Removing the cosmological constant  (setting $\Lambda = 0$) will result in $g(r)$ being
asymptotically $\propto L^2/r^2$, rendering the spacetime asymptotically flat.
Again, we also require that $k=1$ in order to obtain asymptotically Minkowski space with
our solution (\ref{metQ0}).
We have plotted our solution in Figure \ref{flatspace} as a class of solutions of a mass of $M=0.8$.

Similar to the dS situation, we notice the return to a two-horizon
solution when a negative quasi-topological parameter is introduced, without
charge.
We also notice that for $Q=0.6$, the solution is
a naked singularity for Einsteinian, Gauss-Bonnet, and quasi-topological solutions, and the negative quasi-topological
term acts in the same manner as a positive Gauss-Bonnet term.

\mapleplot{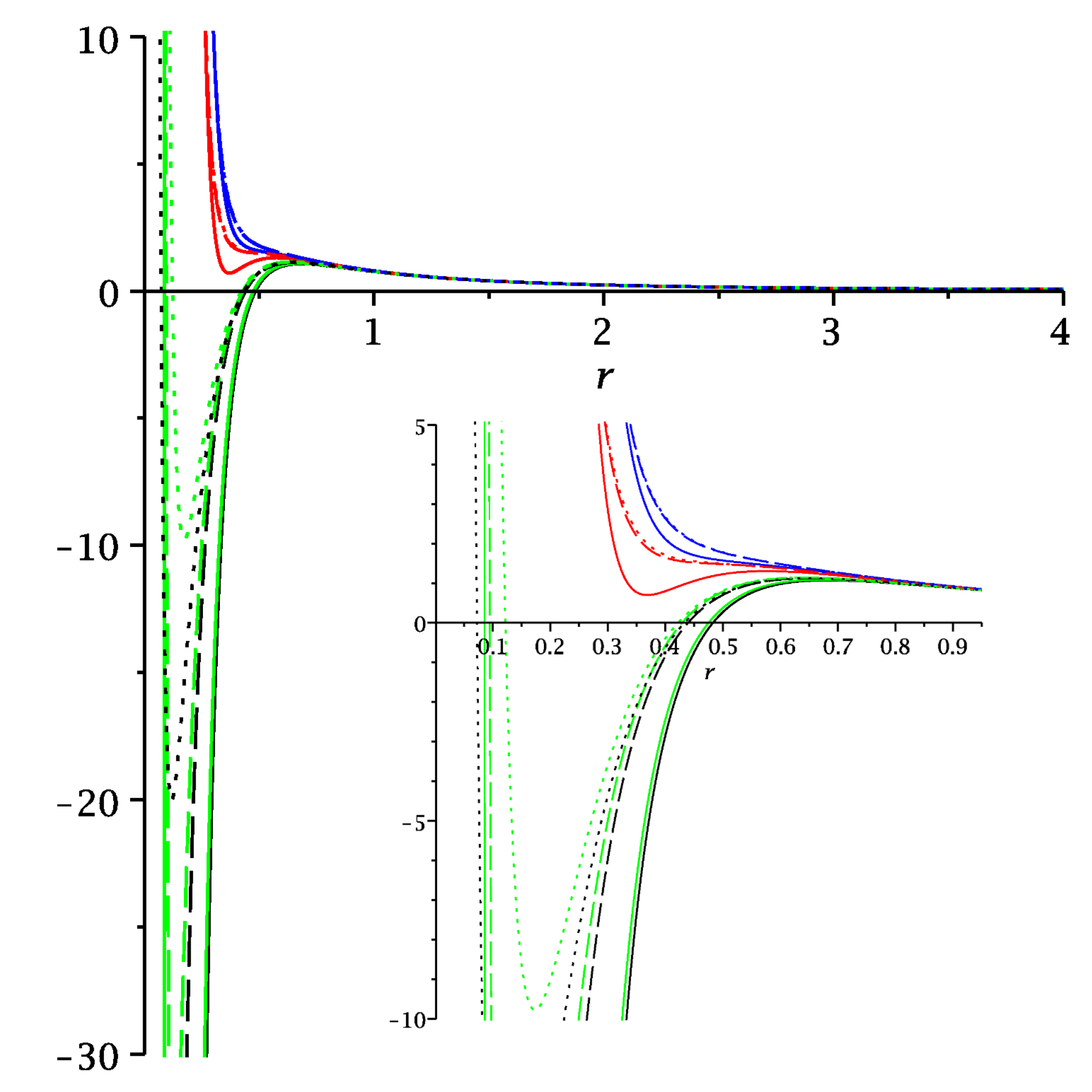}{Solutions in asymptotically flat space for $k=1$, where $Q=0,0.2,0.6,0.7$ are in black, green, red, and blue, respectively. Here $\lambda = 0.04$ and $\mu = -0.001$. Quasi-topological gravity is dotted, Gauss-Bonnet is dashed, and Einsteinian is solid.}{flatspace}{0.4}

\section{RN Thermodynamics}

We compute the entropy and temperature of the Reissner-Nordstr\"om solutions in order
to determine their stability.  
In addition to examining AdS solutions (where $\Lambda = -6 /L^2$),
we have also performed the analysis for asymptotically de Sitter and asymptotically flat solutions, with
$\Lambda = 6/L^2$ and $\Lambda = 0$, respectively.  
The reason we now change our convention and examine a cosmological constant such that the metric functions $g(r)$ do not
converge to the same value in quasi-topological, Gauss-Bonnet, and Einsteinian gravity is because the thermodynamics
are sensitive to the value of the cosmological constant that the metric sees, and we want to study the thermodynamical
behaviour of the black hole solutions under the effect of a single cosmological constant.

We compute the entropy  from the Iyer/Wald prescription 
\cite{iyerwald} where we assume the ansatz
\begin{equation}
S_k = \frac{A}{4 G} \left( 1 + 6 \lambda k \frac{L^2}{{r_h}^2} +%
      9 \mu k^2 \frac{L^4}{{r_h}^4} \right)
\end{equation}
where $D$ is the number of dimensions and $A$ is the surface area of the black hole.
We use this entropy to check the second law of thermodynamics.
The temperature is determined by enforcing regularity of the Euclidean section of the spacetime;
\begin{equation}
T = \left( \frac{r^{2} g^{\prime}}{4 \pi L^{2}}\right)_{r=r_h} 
\end{equation}
where $r_h$ is the radial coordinate of the outermost horizon of the black hole.

Turning next to the first law of thermodynamics for charged quasi-topological black holes,
recall that the first law supplies the equations
\begin{align}
\frac{\partial M}{\partial S_k}_{Q=\text{const}} &= T \\
\frac{\partial M}{\partial Q}_{S=\text{const}} &= \Phi
\end{align}
where $M$ signifies the mass parameter evaluated at horizon (all quantities are at horizon here).
We have verified that this law holds for all of our solutions.  We note that
$\Phi$ is the total electric flux over the horizon, evaluated by
\begin{equation}
\Phi = \int_{S} d^3x A_{r}
\end{equation}
Since curvature doesn't affect the global electrostatic potential (as it's measured at infinity with respect to the horizon),
we are able to use the same result as \cite{hosseinvahidinia}.
That is,
\begin{equation}
\Phi = 16 \pi \frac{Q}{3 {r_h}^2}
\end{equation}

To compute these values we can differentiate the entire expression (\ref{eqnforg}) and solve for $g^{\prime}(r)$.
Substituting values for the mass parameter and charge parameter at the horizon, we are able to absorb leading
coefficients into the values for mass and charge, giving us ``true mass'' and ``true charge'' $m,q$.
We find
\begin{align}
m &= \frac{M}{16 \pi L^2} \\
q &= - \frac{3 Q}{4 \pi}
\end{align}

A negative slope in a temperature-entropy plot produces a situation in which the black hole will be thermodynamically unstable as
its temperature increases with decreasing radius (decreasing entropy), as seen previously for Einsteinian AdS \cite{hawkingpage}.
In the AdS case this causes
small black holes to either decay via Hawking radiation or else tunnel into larger black holes until thermodynamic equilibrium
with the background is reached. However for AdS black holes of sufficiently large charge, we see from
Figure \ref{thermodynamicsads} that no phase transition takes place as the slope is positive in all cases considered. 
Similar to the effect of the quasi-topological term in numerical black hole solutions 
\cite{Brenna:2011gp},   the thermodynamic behaviour of the quasi-topological term closely follows Gauss-Bonnet thermodynamics
for black holes beyond a certain radius (on the order of $r_h \sim 1$). 

\mapleplot{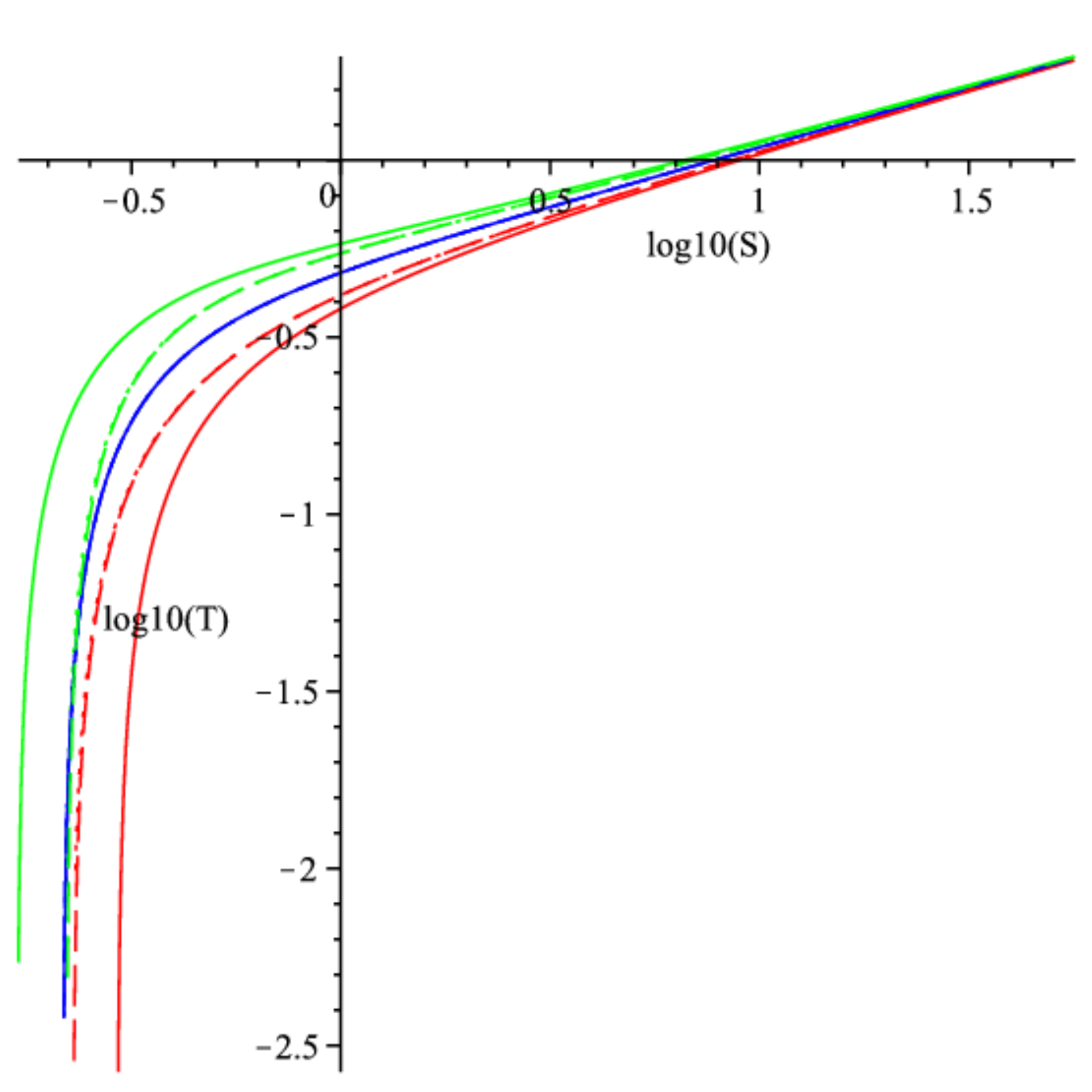}{log(Temperature) versus log(Entropy) for Quasi-topological, Gauss-Bonnet, and Einsteinian (dotted, dashed, solid) asymptotically AdS RN Black Holes of $k=-1,0,1$ (red, blue, green) and $Q=6,\lambda=0.04,\mu=-0.001$;
note that the dotted and dashed lines yield almost identical curves for $k=-1$ and that all three curves are effectively
indistinguishable for $k=0$. 
}{thermodynamicsads}{0.4}
\mapleplot{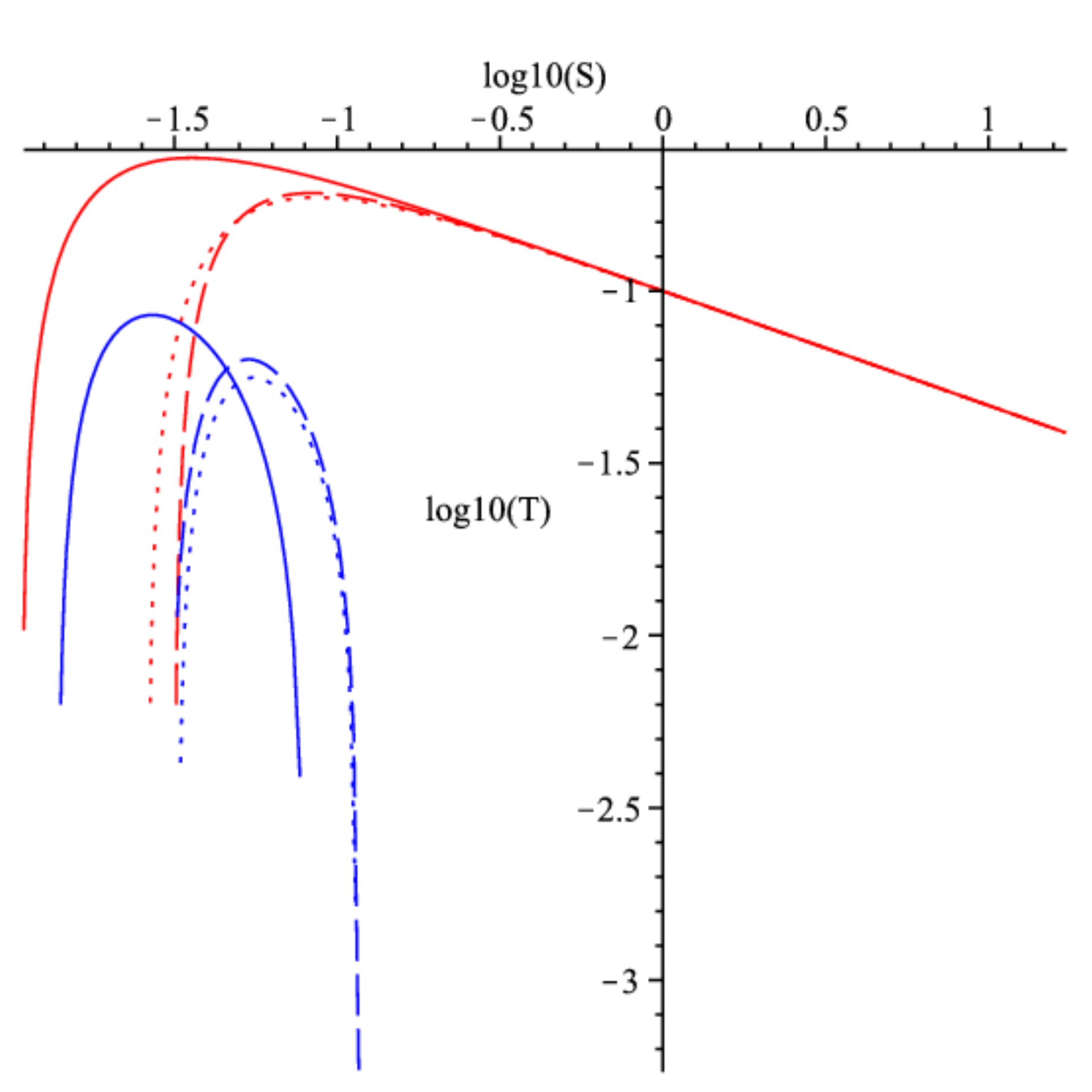}{log(Temperature) versus log(Entropy) for Quasi-topological, Gauss-Bonnet, and Einsteinian (dotted, dashed, solid) asymptotically de Sitter (blue) and asymptotically flat (red) RN Black Holes of $k=1$; $Q=0.6,\lambda=0.04,\mu=-0.001$.}{thermodynamicsds}{0.4}
\mapleplot{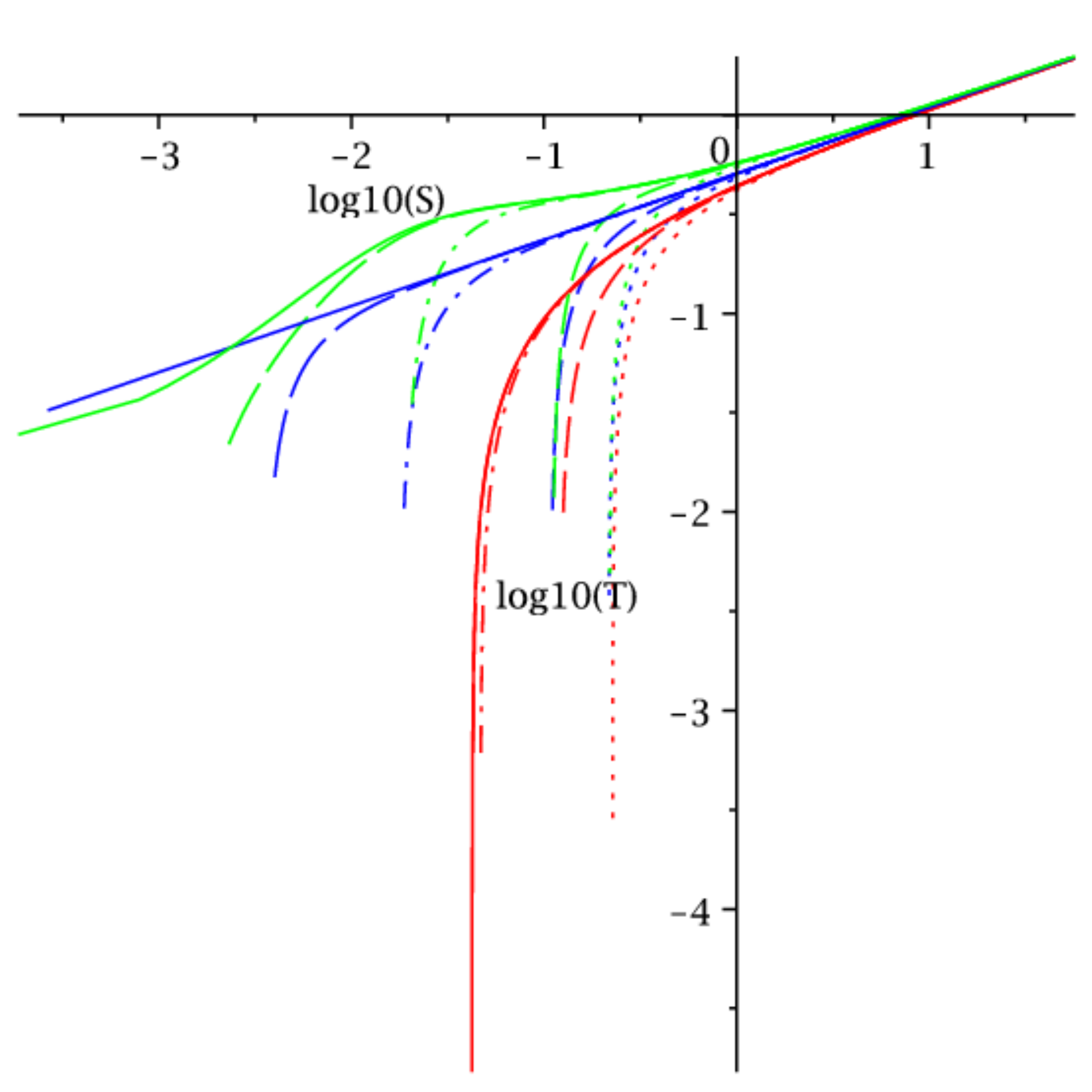}{log(Temperature) versus log(Entropy) for asymptotically anti de Sitter Reissner-Nordstr\"om Black Holes of $Q=6$ (dotted), $Q=3$ (dashed), $Q=0.5$ (dash-dot), $Q=0.1$ (long-dash), and $Q=0$ (solid) for $k=-1,0,1$ (red, blue, green); $\lambda=0.04,\mu=-0.001$. }{effectofQ}{0.4}

In the limit of small black holes, upon closer inspection we see a deviation between the quasi-topological solution and the Gauss-Bonnet solution. We terminate our plots in the limit of extremality for the black holes.
For   $\mu=-0.001$, it is apparent that the quasi-topological term does not induce much change over the Gauss-Bonnet case. We find that
a larger negative value of $\mu$  induces more dramatic behaviour, where for $k=0$ and $k=1$  the slope of the curve in a 
$\log T$ vs. $\log S$ plot asymptotes to horizontal in the limit of small black holes. 
While this behaviour is illustrative of the effect of $\mu$,  it must be taken cautiously because large negative $\mu$ no longer
guarantees positivity of the energy flux in the dual CFT \cite{Myers2}.

We have also plotted solutions for the asymptotically flat and asymptotically de Sitter (Figure \ref{thermodynamicsds}) cases. 
As previously mentioned, only the $k=1$ cases are physical.
Because the de Sitter black holes exist dynamically with a cosmological horizon, an outer horizon, and an inner horizon, we choose
to measure black hole temperature on the outer horizon.
In addition, as the black hole radius of the de Sitter solutions becomes large, the outer horizon approaches the cosmological horizon
(forming a Nariai solution \cite{Cardoso2004}), beyond which the space-time
has a singularity with only a cosmological horizon. The upper bound for the dS case in Figure  \ref{thermodynamicsds} is the
Nariai limit.

In the case of asymptotically flat black holes, we see that our results exhibit close agreement with ref. \cite{Hossein2}.
There, the same behaviour regarding uncharged higher-curvature solutions was found -- not unexpectedly,  as the field equations are
similar. In addition, their thermodynamic analysis showed that the black holes have a region of stability
for $r_{extremal} < r_+ < r_{unstable}$, commensurate with our analysis.
 
We see in Figure \ref{thermodynamicsds}
that the dS black holes undergo a single phase transition from stable to unstable as their horizon radius grows. 
At this transition point (where the curve in Figure \ref{thermodynamicsds} reaches a maximum), the black hole outer radius is $r_+ \approx 0.485$ for the Gauss-Bonnet case. For the flat case, another phase transition
is seen in Figure \ref{thermodynamicsds} for $k=1$,
occurring at the black hole outer radius $r_+ \approx 0.580$ for the Gauss-Bonnet case. 
In the quasi-topological case, these numbers change to  $r_+ \approx 0.496$ and  $r_+ \approx 0.600$, respectively.

In Figure \ref{effectofQ}, we illustrate the effect of the charge parameter $Q$ on the quasi-topological gravity solutions.
The solid lines are uncharged black holes (and exhibit thermodynamic behaviour described in \cite{Brenna:2011gp}). We
see that increasing the charge produces much more stable black holes with larger radii, where a very large increase in temperature is required to deliver a small increase in entropy.
This behaviour is precisely what is expected when increasing the charge - extremal black holes will have larger mass.  
Furthermore, as $Q$ increases the distinction between the different values of $k$ becomes less pronounced, being nearly indistinguishable for
 $Q=6$.

\subsubsection{Specific Heat}

The specific heat is given by  
\begin{equation}
C_v = \frac{\partial M}{\partial T}
\end{equation}
Computing this quantity
\begin{equation}
C_v = \frac{\partial M}{\partial r_+} / \frac{\partial T}{\partial r_+}
\end{equation}
we find that roots exist, and therefore there are unstable regions.
Recall that a negative specific heat will result in a thermodynamic instability with the vacuum \cite{yaffe}.  
For $k=0$, the only real roots of the specific heat are
\begin{equation}\label{cvroots}
r_h = \pm 1/2\,{\frac {\sqrt {2}\sqrt [6]{-{Q}^{2}{\Lambda}^{5}{L}^{4}}}{\Lambda
\,L}}
\end{equation} 
Therefore there are no unstable regions unless $\Lambda$ is negative, as is required in order
to obtain valid black hole solutions for $k=0$; when
$\Lambda = -6$ (AdS), we find that $r_h = \frac{Q^{1/3}}{48^{1/6}}$ is a transition. 
 For  $Q=6$, this gives us a phase transition at $0.953$.

For $k=1$, the equation for specific heat becomes significantly
more complex, and characterizing solutions is much more difficult.
We  begin by looking at the case for $\Lambda = -6$, $k=1$.
Direct substitution leads to the expression 
\begin{equation}
C_v = {\frac { 12 \pi \left( -48\,{r}^{6}-24\,{r}^{4}-24\,\mu+{Q}^{2} \right) 
\left( -{r}^{4}-2\,\lambda\,{r}^{2}+3\,\mu \right) ^{2}}{r
 \left( -288\,{r}^{8}\lambda+720\,{r}^{6}\mu+144\,\mu\,\lambda\,{r}^{2
 }-6\,{Q}^{2}\lambda\,{r}^{2}-48\,{r}^{6}\lambda+336\,{r}^{4}\mu+3\,{Q}
 ^{2}\mu-48\,{r}^{10}+24\,{r}^{8}-72\,{\mu}^{2}-5\,{Q}^{2}{r}^{4}
  \right) }}
\end{equation}
From the numerator, the only non-saddle zeroes will result from the vanishing of the term
$ \left( Q^2 -48 r_h^6 - 24 \mu - 24 r_h^4 \right)$.
We see that a transition therefore occurs under the condition
\begin{equation}
\label{HPtrans}
Q = \pm 2 \sqrt{ 12 r_h^6 + 6 r_h^4 + 6 \mu }
\end{equation}
where $r_h$ is the radius of the transition, denoting $r_+$ by $r_h$ when it corresponds a phase transition radius. 

The denominator produces a tenth-order polynomial whose roots  correspond to divergences
of the specific heat, resulting in complicated phase transition behaviour. 
This type of divergence also occurs in, for example, studies of quantum gravity \cite{husain}.
While explicit (numerical) computation of these roots is a challenging task, we can see that
 roots exist for the value of $Q$ such that  
\begin{equation}
Q = \pm   2 \sqrt {6}  \sqrt { \frac{\left( 12\,{r_h}^{8}\lambda-30\,{r_h}^{6}\mu-6\,\mu\,\lambda
\,{r_h}^{2}+2\,{r_h}^{10}+2\,{r_h}^{6}\lambda-14\,{r_h}^{4}\mu-{r_h}^{8}+3\,{\mu
}^{2} \right)}{\left( 3\,\mu-6\,\lambda\,{r_h}^{2}-5\,{r_h}^
{4} \right)} }
\end{equation}
We need to make sure that this corresponds to a realistic (i.e. real) value of $Q$. For the 
values of $\mu<0$ and $\lambda >0$ in the stable ghost-free region 
 that we consider, the denominator under the square root is negative
and so we must find values for $r_h$ such that the numerator under the square root is also negative.  It is clear
that for sufficiently small $|\lambda|$ and $|\mu|$ that this is always possible -- for example $\lambda=0.015$ and $\mu = 
-0.0005$ there exist real solutions for $Q$. However for $\lambda=0.04$ and $\mu = 
-0.001$ the numerator never becomes positive,
and therefore, there are no divergences in the specific heat.

As another example, consider a cosmological constant ($\Lambda = -1$) for which
$r_h=0.8031577349$.  This results in $Q=1$, a valid solution.
The specific heat is positive and increasing at $r_+ < r_h$, 
and the black hole undergoes a phase transition where
the specific heat becomes singular at $r_+=r_h$.
As $r_+>r_h$, the specific heat increases (though it stays negative), until $r_+ \approx 0.92$ whereupon
the specific heat begins to decrease again.
Another root can also be found: $r_h = 1.59885311$ also results in $Q=1$, and this is identified with another
transition, where the negative specific heat diverges to $-\infty$ and reappears at $r_+>1.59885311$ with positive
value.
As $Q$ increases beyond a critical point (somewhere between $Q=2$ and $Q=3$), the discontinuities in the specific heat
disappear (there are no longer any real positive roots in the denominator), and we have solutions similar to the $\Lambda = -6$ case.

As a verification that our thermodynamics plots are correct, we have checked that the maxima of our curves correspond to
divergences in the specific heat.   For example, for $Q=6$ there is a transition from  negative specific heat to positive specific heat at $r_+=0.877$ in the AdS quasi-topological case ($k=1$). 
Above that radius, the specific heat stays positive, so  our specific heat results agree with the thermodynamics in Figure \ref{thermodynamicsads}.
This transition is unphysical - it happens to be the extremal radius for the black hole and, not coincidentally,
it is also the radius at which we stop our temperature-entropy plot of Figure \ref{thermodynamicsads}.

Switching to $\Lambda=6$ for de Sitter solutions, the function for specific heat is similar, and we solve for the existence of roots 
for $\mu=-0.001$ and $\lambda = 0.04$ in exactly the same manner as for AdS solutions.
At $Q=0.6$, we find that the specific heat crosses from negative to positive for these values of $(\mu,\lambda)$
at $r_+ \approx 0.39$, followed by a divergence to $\infty$ at  $r_+ \approx 0.4945$.
This implies three phase transitions.  The first and third of these
`transitions' are where the outer horizon respectively merges with the inner and cosmological horizons;
both are left off of the plot.
The meaningful phase transition is the one for which the black hole transits from instability (where the temperature (and size) decreases with an increase in entropy)
and a stable phase, where the black hole's temperature increases with an increase in entropy.   The single phase transition can be seen in Figure \ref{thermodynamicsds}  where the curve reaches a maximum.  
Similar behaviour occurs in the Gauss-Bonnet and Einstein
cases for correspondingly smaller values of $r_+$, as is clear from Figure \ref{thermodynamicsds}.
This type of discontinuity in the specific heat has also been studied for Born-Infeld electrodynamics \cite{Banerjee2012}.

Finally, flat space ($\Lambda=0$) solutions also lead to similar occurrences.
For $Q=0.6$,  $\mu=-0.001$, and $\lambda = 0.04$, the specific heat will cross from negative to positive around $r \approx 0.35$,
followed by a divergence to $\infty$ at around $r \approx 0.59$.
The specific heat will then increase from $-\infty$, eventually reaching a turning point and decreasing (never crossing zero) .  
This implies two phase transitions.  Again, the first transition is where the black hole becomes extremal and is left off of the plot.  
The remaining phase transition is again the black hole's transition from an unstable phase (where they possess large horizon radii)
to a stable phase (for black holes with small horizon radii).
This is again seen in our temperature/entropy plot of Figure \ref{thermodynamicsads}.

Summarizing,
 the intriguing result is that we effectively have control of a $5^{th}$ order polynomial with coefficients that depend
on $\Lambda$, $\mu$, $\lambda$, and $Q$, allowing us to theoretically (without constraints on $\mu$ or $\lambda$)
produce up to five discontinuities in the specific heat.
In general, this polynomial will contain five important coefficients and one scale factor (which is unimportant when computing its zeroes).
In the Gauss-Bonnet scenario we have three parameters upon which four coefficients in the polynomial depend, before $\Lambda$ is fixed, as the constant term disappears
(group the denominator of the specific heat into terms of $r^2,r^4,r^6$, etc.).
For the quasi-topological case we have four parameters to fix five coefficients.
An additional quasi-topological term \cite{Dehghani:2011vu} might allow us to obtain the final parameter, and perhaps it will form an independent
basis of these parameters on each term, giving us access to a maximum of five phase transitions (that is, allowing us to redefine four
unique parameters $\alpha_1,\alpha_2,\alpha_3,\alpha_4,\alpha_5$ for which the denominator of the specific heat appears as 
$r^{10} + \alpha_1 r^8 + \alpha_2 r^6 + \alpha_3 r^4 +\alpha_4 r^2 + \alpha_5$).  
 
While adding a fourth-order quasi-topological term appears to be a useful endeavour, we leave its effect open to question since
it is not immediately clear, if successful, what utility a complete basis for the roots would be, given that $^{\text{a)}}$ the 
solution to the quintic is not in general expressible in radicals and $^{\text{b)}}$ the parameters $\mu$, $\lambda$, $Q$ all
have various restrictions and so they do not extend over the reals.
In addition, the possibility exists that a quartic quasi-topological term will increase the order of the polynomial (though we do not
expect this to be the case, as it did not increase from Gauss-Bonnet to third order quasi-topological gravity).

\section{Charged Black Holes in $K$-th order Quasi-topological Gravity}

In this section we consider $K$-th order quasi-topological gravity in $(N+1)$ dimensions.  The specific form of the Lagrangian for $K$-th order quasi-topological gravity has only been obtained for $K=3$ \cite{Myers} and $K=4$ \cite{RBMMHD}. 
However we can make some remarks for general $K$ based on this work,  even though 
the specific Lagrangian has yet not been found. 

Consider the spherically symmetric ansatz
\begin{equation}\label{metric2}
ds^2 = -\frac{r^2}{L^2} N^2(r) g(r) dt^2 + \frac{L^2 dr^2}{r^2 g(r)} + \frac{ r^2 }{ L^2 } d\Sigma_k^2
\end{equation}
\begin{equation}\label{gfield2}
A  = h(r) dt 
\end{equation}
where $d\Sigma_k^2$ is the metric of a $(N-1)$-dimensional constant curvature hypersurface.

Maxwell's equations yield 
\begin{equation}
\left( \frac{r^{N-1}}{N} \left(h \right)^\prime  \right)^\prime = 0
\end{equation}
implying
\begin{equation}\label{gfield3}
A  = \frac{Q}{(N-2) r^{N-2}} dt 
\end{equation}
as the solution for the gauge field.
Since the field equations can only be $2^{nd}$ order in the metric functions in the spherically symmetric case, it
is reasonable to conjecture that the field equation for the metric function $g(r)$ is
\begin{equation}
 \left(r^{N} \sum_{k=0}^K \mu_k \kappa^k \right)^\prime = \frac{ r^{N-1}}{2}\left[  \left( \frac{Q}{r^{N-1}}\right)^2  \right]  \label{genkap}
\end{equation}
for $K$-th order quasi-topological gravity in $(N+1)$ dimensions, where $\kappa=(k L^{2}r^{-2}-g)$.  We easily obtain
\begin{equation}
\Upsilon[g] = \sum_{i=0}^K \mu_i \kappa^i = \frac{M}{r^N} -
 \frac{Q^2} {(N-2) r^{2N-2}} 
  \label{genkap2}
\end{equation}
and the metric function for Einsteinian gravity (setting $\mu_1 = 1$) is
\begin{equation}
\label{planarsol}
\frac{r^2}{L^2} g(r) =   \frac{\mu_0 r^2}{L^2}  + k - \frac{M}{L^2 r^{N-2}} +
 \frac{Q^2} {L^2 (N-2) r^{2N-4}} 
\end{equation}
This equation has the same form as the corresponding $Q=0$ situation in Lovelock gravity \cite{Camanho:2011rj,Soda}, where
 $K \leq \left[ \frac{n}{2} \right] $.   In quasi-topological gravity $K$ is not restricted (though there are certain values of $N$ for which the construction does not make sense).

We start by considering the case where $k=0$.
We find that the planar solution of equation (\ref{planarsol}) admits a vanishing $g(r)$ only when
\begin{equation}
- \mu_0  + \frac{M}{{r_h}^N} - \frac{Q^2}{\left( N - 2\right) {r_h}^{2 N - 2}} = 0
\end{equation}
or $r_h$ corresponding to roots of
\begin{equation}
\label{rroots}
M {r_h}^{N-2} - \mu_0 {r_h}^{2N-2} - \frac{Q^2}{N-2}
\end{equation}
These roots can be numerically obtained.
We can see that in higher dimensions, the high polynomial degree
results in a rich dependency of horizon solutions on $M$ and $Q$.

Now, more generally (for any $k$), at the horizon we have
\begin{equation}
\kappa = \frac{k L^2}{{r_h}^2}
\end{equation}
so that 
\begin{equation}
\Upsilon \left[\frac{k}{{r_h}^2} \right] = M {r_h}^{N-2} - \mu_0 {r_h}^{2N-2} - \frac{Q^2}{N-2}
\end{equation}
This is not analytically solvable but the branch we are
interested in should be a perturbation of the branch for the uncharged case \cite{Camanho:2011rj}.
Our solution will be a continuous transform of this case
and so we also relate our analysis to the Einstein-Hilbert (EH) branch.

For a sufficiently small value of $Q$, our Gauss-Bonnet EH-branch is
\begin{equation}
g = \frac{1}{2 \lambda} \left( 1 + \frac{2 \lambda k L^2}{r^2} -
\sqrt{ 1 - 4 \lambda \left( 1 - \frac{M}{3 r^4} + \frac{Q^2}{24 r^6} \right)}
\right)
\end{equation}
We can see that this is equivalent to our previously obtained solution for $g(r)$ as given
in Appendix \ref{asolnomu}.
This also happens to be the branch of the solution connected to the Einsteinian solution;
as $\lambda \rightarrow 0$, we recover the Einsteinian solution (\ref{einstsoln}).

\section{Discussion}

In this paper we have examined Reissner-Nordstr\"om-type black holes in quasi-topological gravity.
We have introduced third-order quasi-topological curvature terms to the Lagrangian, and we find that
the introduction of a Maxwell field leads to AdS solutions which have the double-horizon characteristic
of the Reissner-Nordstr\"om black hole.
As expected, the small quasi-topological terms do not drastically alter the solutions.

We have seen that not only can the Maxwell charge parameter induce two horizons in flat and de Sitter spacetimes (as expected), but a quasi-topological parameter -- within the stable ghost-free region of parameter space -- 
is able to create two horizons without any charge parameter present.
In addition, along with the Gauss-Bonnet parameter, the higher curvature correction terms are able to take a near-extremal black hole
in Einstein gravity into a singularity with no horizons, in flat space.
 
We have also provided further verification that a small, negative quasi-topological term will act similarly to the
effect of a positive Gauss-Bonnet term with respect to stability.
We have also seen how the quasi-topological parameter can affect phase transitions.
Because the quasi-topological parameter is restricted to small values, we are not able to add or remove phase transitions
at will, but we do see that for specific cases, $\mu$ is able to broaden the region of thermodynamical parameter space
that we are able to probe.
Specifically, the specific heat divergences are governed by a quintic polynomial in black hole radius.
The Gauss-Bonnet solution allows us to control two coefficients independently, while the quasi-topological
term gives us control over three of the five coefficients (once the cosmological constant had been determined).

Regarding future work, Tarrio and Vandoren \cite{Tarrio} have shown parallels to Reissner-Nordstr\"om black holes
in gravity with Lifshitz scaling symmetries, with unexpected behaviour.
It would be interesting to see the modifications to these behaviours in a Lifshitz symmetric
gravitational theory \cite{Dan,Deh} with higher order quasi-topological terms.
Another interesting phenomenon is the surprising result of instability of RN black holes in
dimensions greater than 6 \cite{Cardoso}.
Will this result hold in the quasi-topological regime?

\appendix
\section{Analytic Asymptotically Anti de Sitter Solutions for $M$ given $r_+$}
\label{asolnomu}

For Gauss-Bonnet gravity, we obtain a simplified solution for $f(r)=g(r)$ of
\begin{equation}
g(r) = \frac{6 r^3 + 12 \lambda k L^2 r - \sqrt{36 r^6 - 144\lambda r^6 - 6\lambda L^8 Q^2 %
	+ 48 \lambda M r^2}}{12 \lambda r^3}
\end{equation}
and, setting $L=1$ and solving roots for $M$, we see that
\begin{equation}
M = \frac{-24 {r_+}^6 - 24 \lambda {r_+}^6 + Q^2 + 24 {r_+}^4 k + 24 \lambda k^2 {r_+}^2}{8 {r_+}^2}
\end{equation}

Similarly in Einsteinian gravity, 
\begin{equation}
\label{einstsoln}
g(r) = \frac{L^8 Q^2 - 8 M r^2 + 24r^6 + 24 k L^2 r^4}{24 r^6}
\end{equation}
with roots
\begin{equation}
M = \frac{Q^2 + 24 {r_+}^6 + 24{r_+}^4 k}{8 {r_+}^2}
\end{equation}

\acknowledgments

This work was supported in part by the National Sciences and Engineering Research Council of Canada.

\end{document}